\documentclass[a4paper,10pt,twoside]{cpc-hepnp}

\usepackage{multicol}
\usepackage{graphicx}
\usepackage{booktabs}
\usepackage{amssymb,bm,mathrsfs,bbm,amscd}
\usepackage[tbtags]{amsmath}
\usepackage{lastpage}
\usepackage{CJK}
\usepackage{pifont}

\begin{document}
\begin{CJK*}{UTF8}{gbsn}

\fancyhead[c]{\small Chinese Physics C~~~Vol. XX, No. X (2015)
XXXXXX} \fancyfoot[C]{\small 010201-\thepage}

\footnotetext[0]{Received xx March 2015}

\title{Test of a fine pitch SOI pixel detector with laser beam\thanks{Supported by National Natural Science
Foundation of China (11375226) }}

\author{%
      LIU Yi(刘义)$^{1,2,3;1)}$\email{liuyi@ihep.ac.cn}%
      \quad LU Yunpeng(卢云鹏)$^{1,2;2)}$\email{yplu@ihep.ac.cn}%
      \quad JU Xudong(鞠旭东)$^{1,2,3)}$
      \quad OUYANG Qun(欧阳群)$^{1,2)}$
}
\maketitle

\address{%

  $^1$ State Key Laboratory of Particle Detection and Electronics, Beijing 100049, China\\
  $^2$ Institute of High Energy Physics, Chinese Academy of Sciences, Beijing 100049, China\\
  $^3$ University of Chinese Academy of Sciences, Beijing 100049, China\\
}

\begin{abstract}

  A silicon pixel detector with fine pitch size of 19$\times$19 $\mu$m, developed base on SOI (silicon on insulator) technology, was tested under the illumination of infrared laser pulses. As an alternative way to particle beam tests, the laser pulses were tuned to very short duration and small transverse profile to simulate the tracks of MIPs (minimum ionization particles) in silicon. Hit cluster sizes were measured with focused laser pulses propagating through the SOI detector perpendicular to its surface and most of the induced charge was found to be collected inside the seed pixel. For the first time, the signal amplitude as a function of the applied bias voltage was measured for this SOI detector, deepening understanding of its depletion characteristics.

\end{abstract}

\begin{keyword}
silicon pixel detector, SOI, infrared laser
\end{keyword}

\begin{pacs}
29.40.Cs, 29.40.Gx
\end{pacs}

\footnotetext[0]{\hspace*{-3mm}\raisebox{0.3ex}{$\scriptstyle\copyright$}2013
Chinese Physical Society and the Institute of High Energy Physics
of the Chinese Academy of Sciences and the Institute
of Modern Physics of the Chinese Academy of Sciences and IOP Publishing Ltd}%

\begin{multicols}{2}

\section{Introduction}
With the increasing requirements of modern particle physics experiments, the design and construction of sophisticated detector systems is ever more challenging. The innermost vertex detectors for the next generation electron-positron collider experiments are among the most demanding detectors and required to be with high spatial resolution but very low material budget. To achieve a spatial resolution of 5 \(\mu\)m, which is necessary for secondary vertex measurement at the International Linear Collider (ILC) experiments\cite{lab_aa}, pixel detectors with fine pitch of 17 \(\mu\)m need to be explored. The material budget is limited to $\sim$0.15\% \(X_{0}\) per detector layer to reduce the multiple scattering effects. The SOI (silicon on insulator) pixel detector, with the clear advantages of high integration and fine pitch size, represents a very promising candidate for such pixel detectors. A prototype SOI pixel detector with pitch size of 19 $\mu$m has been developed and its main characteristics are reported below.

Minimum ionization particles (MIPs) generated with high energy accelerator are typically used to characterize pixel detector. They can provide precise spatial and time information of particles propagating through the detectors. However it is not always convenient due to limited accessibility and availability of test beam facilities. \(\beta\)-rays from radioactive sources are also frequently used in laboratory tests. However the electron energy spectrum is broad and low energy electrons tend to deposit more energy than that of MIPS along the particle trajectory. Therefore they cannot reproduce the same effects of MIPs. As an alternative technique, focused infrared laser pulses are commonly used in laboratories to simulate MIPs effects. In addition, the position and time of the incident pulses are controllable, which makes the infrared laser instrument easy to operate for detector tests.

It was reported in Refs.~\cite{lab_yu,lab_kh} that a ND-YAG laser system was used to measure the charge sharing between nearby strips in the silicon micro-strip detectors for the ATLAS experiment. Inspired by such measurements, a compact system based on a low-power semiconductor laser device has been developed for characterize the SOI pixel detector. It should be pointed out that for the silicon micro-strip detectors tested in Refs~\cite{lab_yu,lab_kh}, only a small fraction of the infrared photons were reflected or blocked by the aluminium on top of the detector surface. For silicon pixel detectors, however, the surfaces are typically covered by the aluminium path of the readout circuits or metal dummy\cite{lab_fo}, which can prevent almost completely the infrared photons penetrating through the detector bulk. To overcome such effects, the SOI detector was designed with a centering opening on the surface of each individual pixel. During the tests, the laser pulses were focused to a few micrometers and well through the aluminium openings. With the proper setup of the laser system, measurements of the hit cluster profile and depletion characteristics of this prototype SOI detector were performed. Results on the size of the transverse diffusion of the stimulated tracks as well as the development depth in the sensitive volume were obtained.

\section{SOI pixel detector and readout system}

The SOI detector were fabricated with two active silicon layers, the high resistive detector substrate and the thin top device layer. They were separated by an electrically isolating buried oxide layer (BOX), as depicted in Fig.~\ref{fig_soi_cross}. A high dose of boron was implanted to the top surface of the substrate, by which a matrix of diodes was formed. Holes were etched through the BOX layer to allow electrical contact between the anode of each diode and the pixel circuit right on top. There were in total five metal layers used for high-density circuit interconnection and power distribution on the top of the device layer. The whole pixel area was covered by metal layers except for the rectangle opening on top of each pixel, which was designed for laser illumination. The backside of the detector was covered entirely by aluminium, which could serve as a cathode shared by all the pixels. The depleted silicon volume extended from the pixel side to the backside as the applied bias voltage increased. The substrate with a thickness of 260 \(\mu\)m was expected be completely depleted with a bias voltage of about 200~V~\cite{lab_yl}.

\begin{center}
\includegraphics[width=8cm]{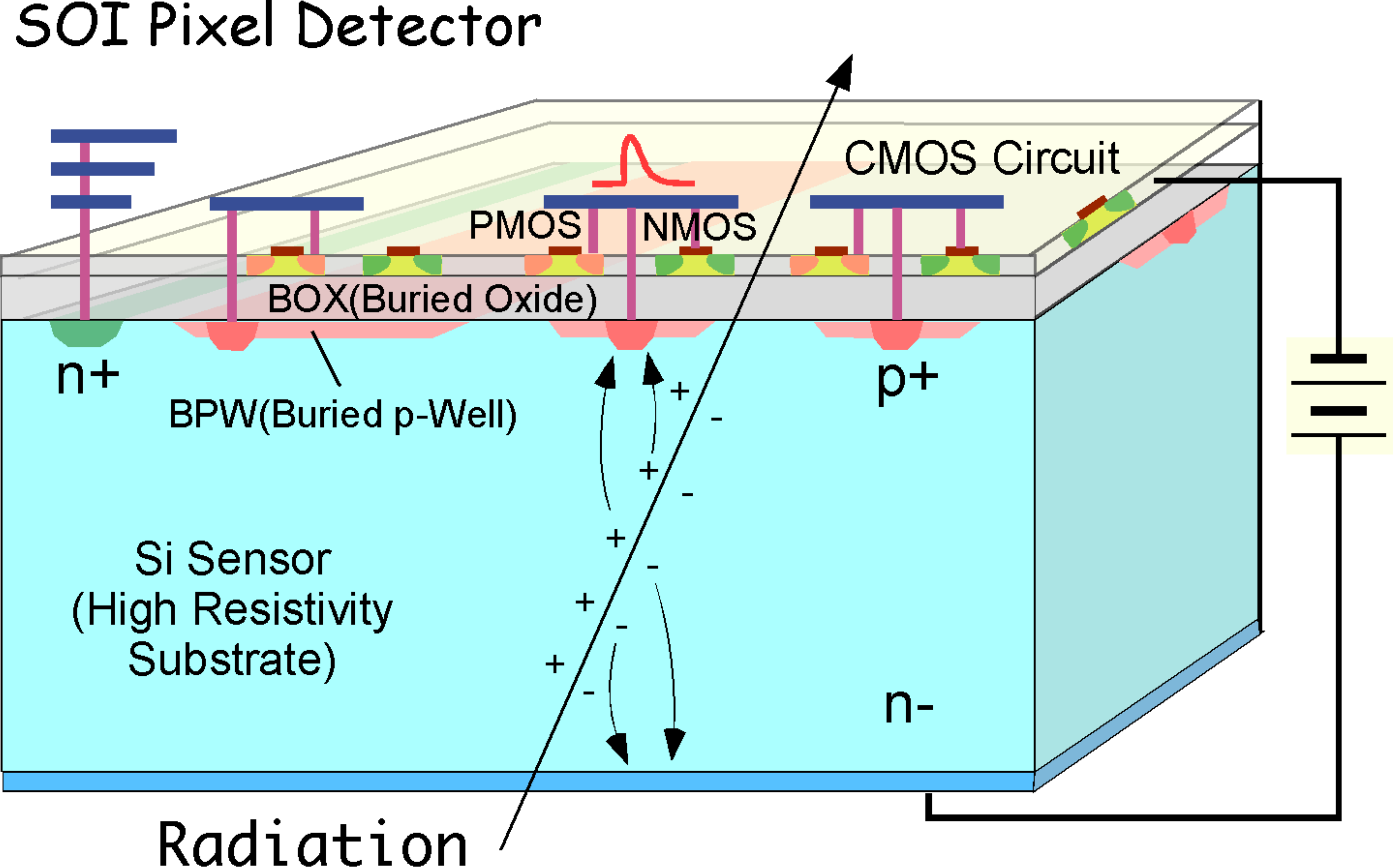}
\figcaption{\label{fig_soi_cross} Schematics of SOI pixel detector, taken from Ref\cite{lab_ya}.}
\end{center}

Integrated circuits were fabricated with the standard 0.2 \(\mu\)m CMOS process.  There were a total of 64\(\times\)64 pixels with pixel size of 19\(\times\)19 \(\mu\)m\(^2\), covering 1.3\(\times\)1.3 mm\(^2\) of the detector area. The remaining insensitive area of 3\(\times\)3 mm\(^2\) was occupied by the peripheral circuit and the I/O pads. The stimulated charge in the detector substrate were driven by the electric field to the nearest pixel electrode and integrated by the input capacitance of the front-end circuit, which was then converted to an analog voltage. Pixel signals were read out in series under the control of the peripheral circuit and driven by an analog buffer circuit, which was shared by all the pixels, to the a separate commercial ADC chip on the test board. The readout time of an entire frame was about 500 \(\mu\)s.

This detector die was ceramic-packaged with 176 pins, with an encasement featuring a removable lid to make the upper surface of the die visible from outside. Laser light could illuminate the die and penetrate through both the circuit and BOX layers and reach the detection substrate. The packaged detector was mounted on an adapter PCB board, which itself was mounted onto a FPGA based data acquisition system board named SEABAS. There were several features provided by the SEABAS board, including timing control, power supply, AD/DA conversion, and 100M Ethernet link. A dedicated program was developed based on C++ and the ROOT framework to configure, monitor and collect data from the detector under test via the Ethernet port.

\section{Laser test setup}

The laser system consisted of a laser generator, an optical fiber, a collimator, a focusing lens, and a 3D linear motion platform (Fig.~\ref{fig_system_setup}). In order to simulate charged particles imprinting the silicon detectors, laser pulses generated from a picosecond infrared semiconductor laser generator were focused to tiny spot and tuned to short time burst. The pulse power were tunable to make the energy deposit equivalent to that of MIPs. For laser with a wavelength of 1064~nm (\textit{i.e.} 1.17~eV per photon), its attenuation length in crystal silicon was estimated to be 1036 \(\mu\)m\cite{lab_ed}, which allowed the intensity of laser beam almost constant down to a depth 260 \(\mu\)m in the substrate. There would be about 20000 e/h pairs generated by a MIP penetrating through 260 \(\mu\)m thick silicon. To generate the same amount of charge, laser energy of 3.5 fJ should be deposited in the sensitive volume. However, taking into account the optical transmission loss and the quantum absorption efficiency, the initial power from the laser generator should be one order of magnitude higher than the absorbed power. The duration of the laser pulse could reach a minimum close to 100 ps, which would be much shorter than the charge collection time (a few ns). In addition, the laser pulse generation was synchronized with the readout electronics. This allowed an exact number of laser pulses to be injected to the sensor between two consecutive readout operations.

\begin{center}
\includegraphics[width=8cm]{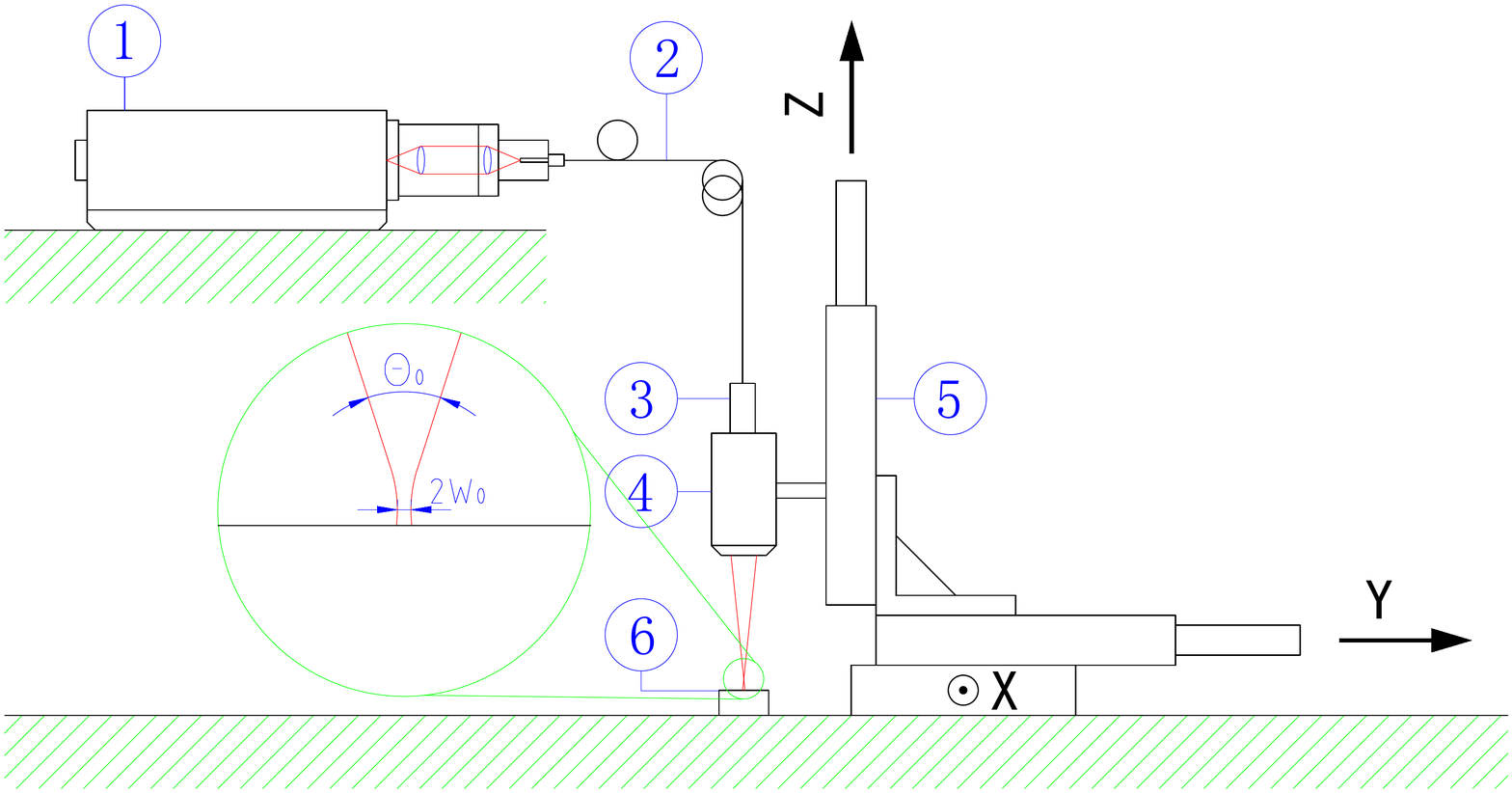}
\figcaption{\label{fig_system_setup} Laser test setup: \ding{172} laser generator, \ding{173} optical fiber, \ding{174} collimator, \ding{175} lens, \ding{176} 3D linear motion platform, \ding{177} detector chip. }
\end{center}

The laser light was guided to a collimator through a single mode fiber and focused by a lens. The relationship among aperture \(D\), focal length \(F\) and divergence angle \(\theta_0\) could be formulated as:
\begin{equation}
\label{eq_tan}
\tan\theta_0 = \frac{D}{F}
\end{equation}

With a focal length of 10 mm and an aperture of 4 mm, a divergence angle would be 23\(^\circ\) according to Eq.~\ref{eq_tan}. For a Gaussian laser beam with a wavelength of \(\lambda\), the beam waist diameter \(2\omega_0\) (namely the minimum spot size that can be achieved) could be calculated as:
\begin{equation}
\label{eq_w0}
2\omega_0 = \frac{4\lambda}{\pi\theta_0} \approx \frac{4\lambda}{\pi}\frac{F}{D} 
\end{equation}

Eq.~\ref{eq_w0} suggested that \(2\omega_0\) and \(\theta_0\) could not decrease at same time for certain wavelength, which would be a critical limiting factor for laser tests. The beam waist diameter \(2\omega_0\) would be $\sim$3.4 \(\mu\)m for the current setup.

The collimator and lens were mounted on a motor-driven platform with programmable \(XYZ\) coordinates. It could provide a 25 mm range in each direction that covered the full area of the SOI detector under test. Absolute position precision could reach 1.5 \(\mu\)m in each direction. To prevent the vibration from laboratory surrounding, the entire system was mounted on a stable optical bench. Platform motion was controlled by sending commands to the serial port of the motor controller.

\section{Measurements}

\subsection{Beam characterization and alignment}

The highly focused and low power infrared laser beam was invisible to the eyes and difficult to measure with an economically affordable instrument. Instead, a special procedure was developed to focus and align precisely the laser beam to a single pixel. It took advantage of the fine pitch size of the SOI detector, which enabled the possibility to image and characterize the laser beam. The procedure consisted the following two steps:

\begin{itemize}
\item
  Scan along the Z axis with the lens to locate the correct Z coordinated, where the beam waist was right in the window plane.

\item
  Scan in the XY plane to locate the central coordinates, where the beam hit the window of the target aluminium opening.

\end{itemize}
  
Each pixel was 19\(\times\)19 \(\mu\)m\(^2\) with a 5\(\times\)9 \(\mu\)m\(^2\) aluminium opening on its upper surface. As the laser pulses could be focused to 3.4 \(\mu\)m at the focal point, the aluminium opening was wide enough to allow the laser beam to reach the detector substrate. However, due to the relatively large divergence angle, the area illuminated by the beam increased as the beam waist deviating from the window plane. The diameter of the laser beam projected on the window plane was measured as a function of focal lens height. Linear fits were performed to determine the height, where the spot size was at its minimum. A divergence angle of 19\(^\circ\) was also extracted from this linear fit, which differed from the calculated result of 21.8\(^\circ\) with Eq.~\ref{eq_tan}. This was largely due to the non-perpendicular incident angle of the laser beam and the shifted center of the illumination spot as the lens height was adjusted. This incident angle was estimated as 2.5\(^\circ\) and labeled as \(\Phi_0\) in Fig.~\ref{fig_laser_shape}. The impacts of divergence angle and incident angle on the cluster size were discussed in Section 4.2.

\begin{center}
\includegraphics[width=8cm]{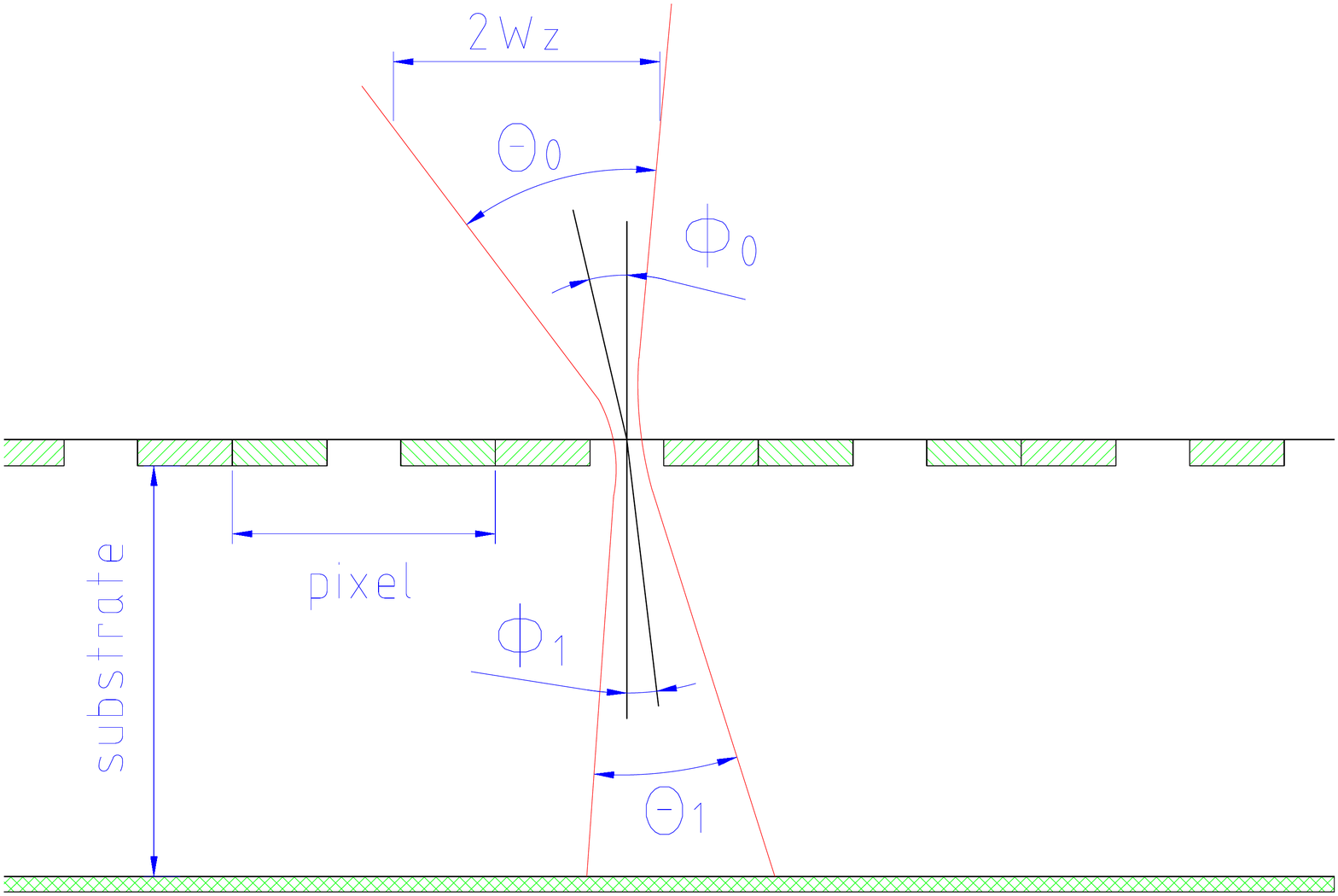}
\figcaption{\label{fig_laser_shape} Laser beam focused on an aluminium opening. \(\theta_0\), the angle of laser divergence in air; \(\theta_1\), the angle of laser divergence in silicon; \(\phi_0\), the angle of laser obliquity in air; \(\phi_1\), the angle of laser obliquity in silicon; \(2w_z\), the diameter of laser beam at a Z position. }
\end{center}

\begin{center}
\includegraphics[width=4.2cm]{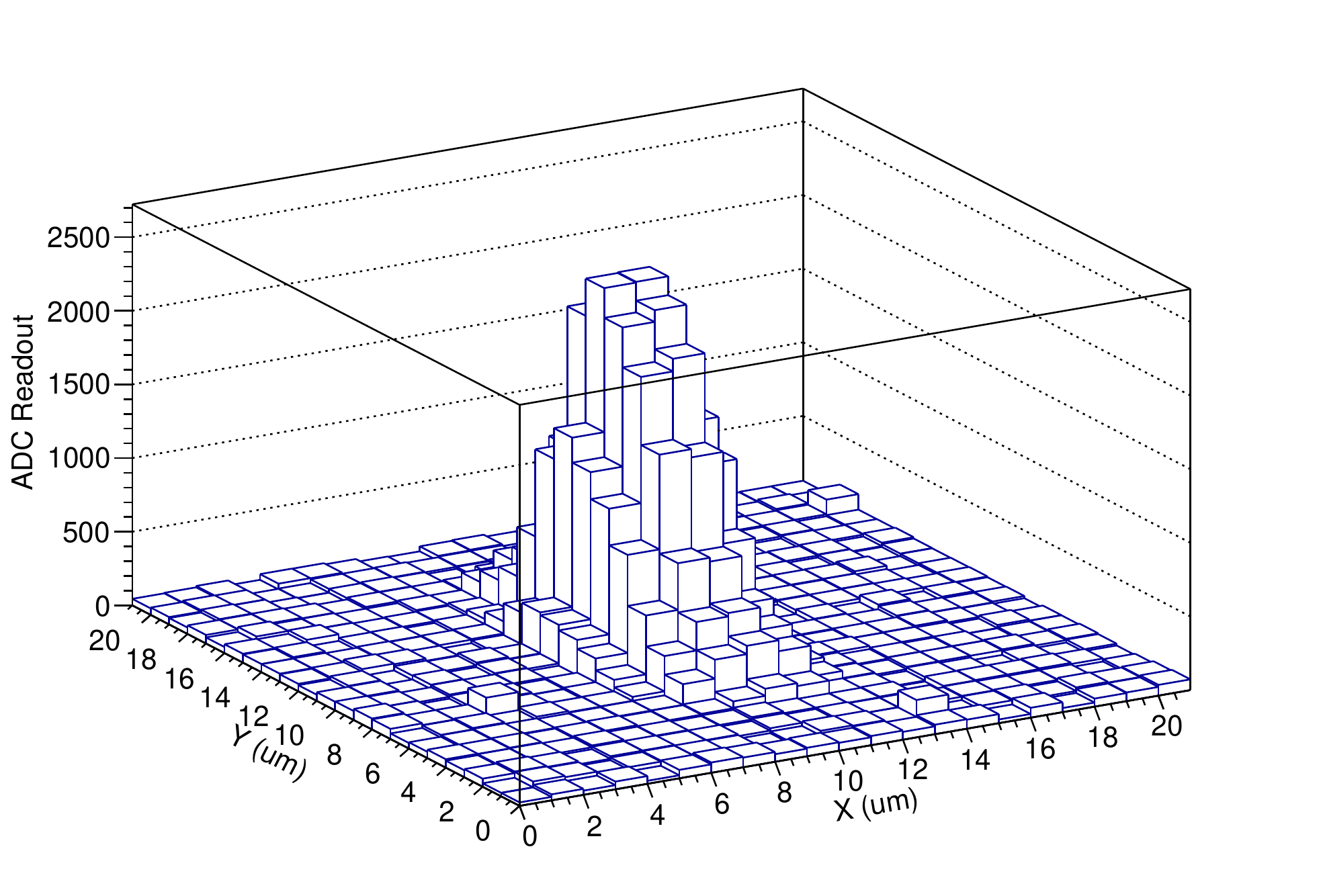}
\includegraphics[width=4.2cm]{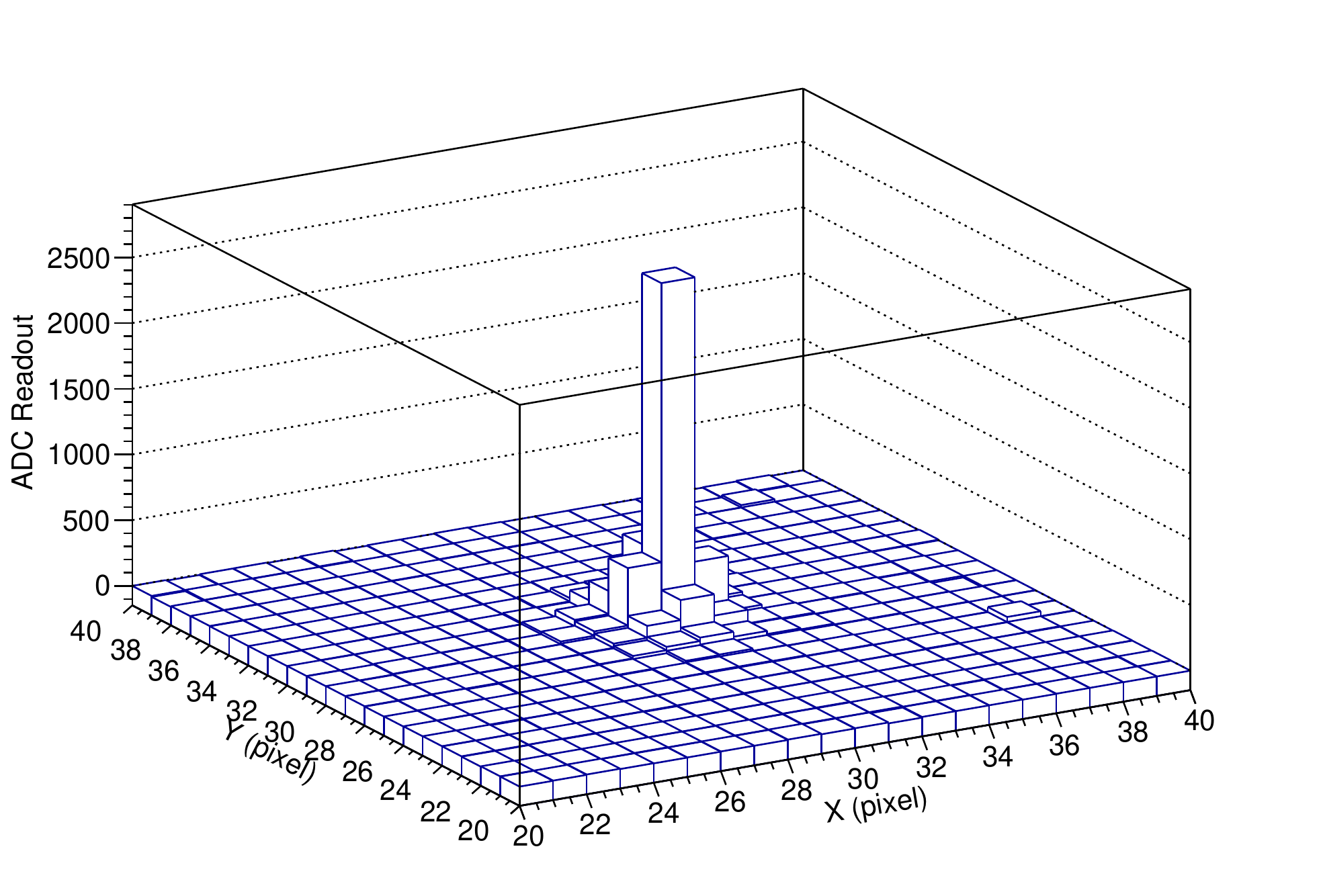}
\figcaption{\label{fig_pixel_response} Histogram of signal amplitude dependence on laser incident position, with a 19\(\times\)19 \(\mu\)m\(^2\) area belonging to the target pixel scanned (left). The hit cluster of a single well focused laser pulse (right).}
\end{center}

After adjusting the laser beam to its focal plane, a two-dimensional scan with a step-size of  \(\mu\)m in each direction was performed to obtain the response map of the SOI detector. As shown in Fig.~\ref{fig_pixel_response} (left), the signal amplitude of the target pixel showed a clear dependence on the laser incident position. The signal amplitude reached the maximum when the laser beam hit the exact center of the opening, and dropped to zero abruptly when the beam moved out of the window because the laser beam was totally blocked by the aluminium covering. In addition, as shown in Fig.~\ref{fig_pixel_response} (right), the seed pixel could be clearly distinguished from its neighboring pixels when it was directly hit by the laser beam. In these studies, the laser beam was not required to deposit the same amount of  energy as that of MIPs.

\subsection{Sensor depletion}

Full depletion as one of the most important characteristics of SOI silicon detectors was also studied. As shown in Fig.~\ref{fig_response_voltage_iv} (left), the signal amplitude of the seed pixel increased with higher applied bias voltage (\(V_{bias}\)) and reached a plateau at around 190 V. The total signal, summing over the charge collected by all the pixels, showed the same trend but with higher amplitude. In the ideal case, the signal amplitude should increase proportional to \(\sqrt{V_{bias}}\) and then saturate at a certain point~\cite{lab_hs} after reaching full depletion. However, due to divergence and reflection, the energy deposition along the laser path in the sensitive detector volume was not constant: such effects could distort the curve but would not shift the turning point. The I-V curve was measured for the same SOI detector and shown in Fig.~\ref{fig_response_voltage_iv} (right). With the turning point, the full depletion voltage could be determined unambiguously.

\begin{center}
\includegraphics[width=4.2cm]{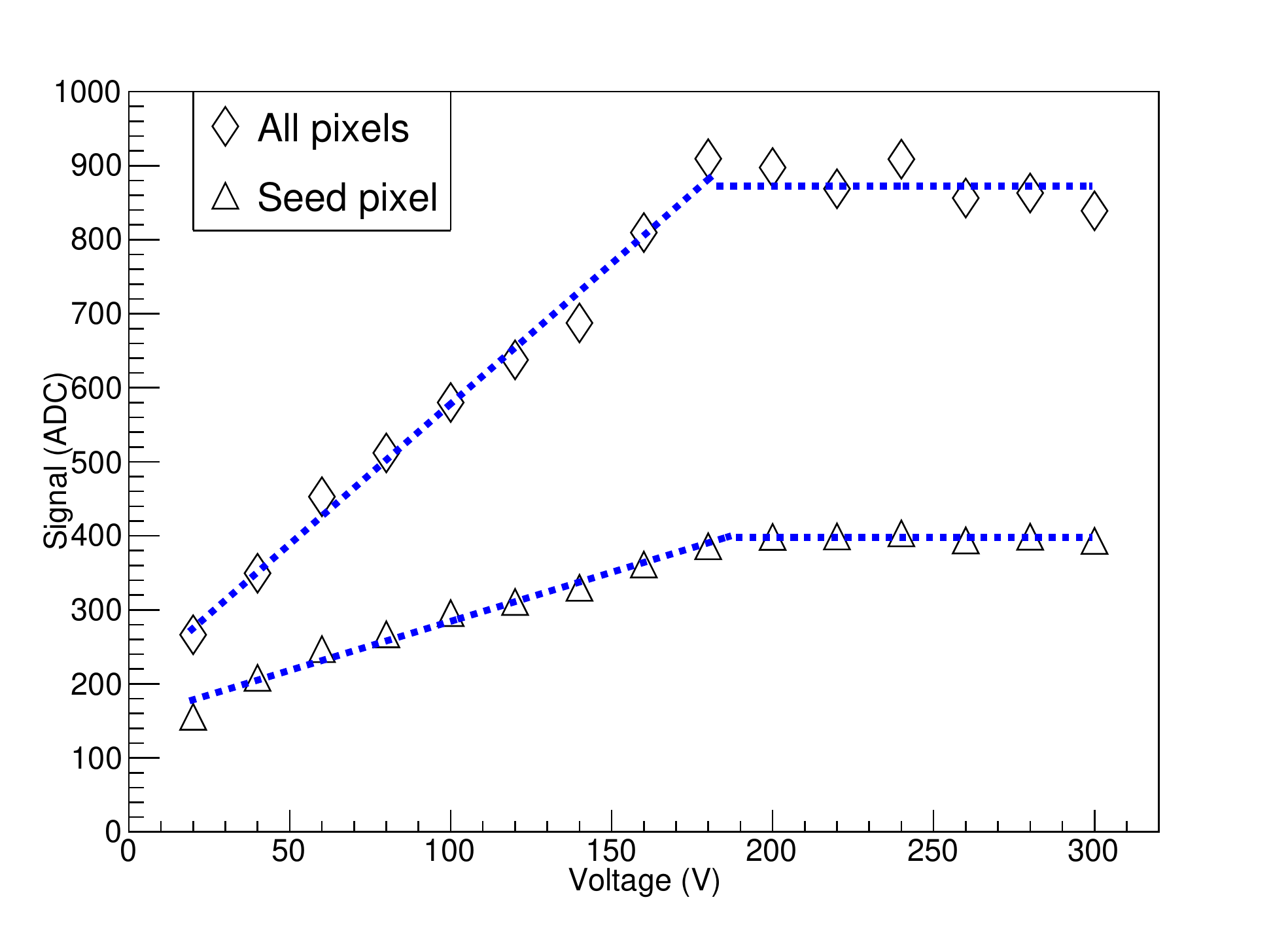}
\includegraphics[width=4.2cm]{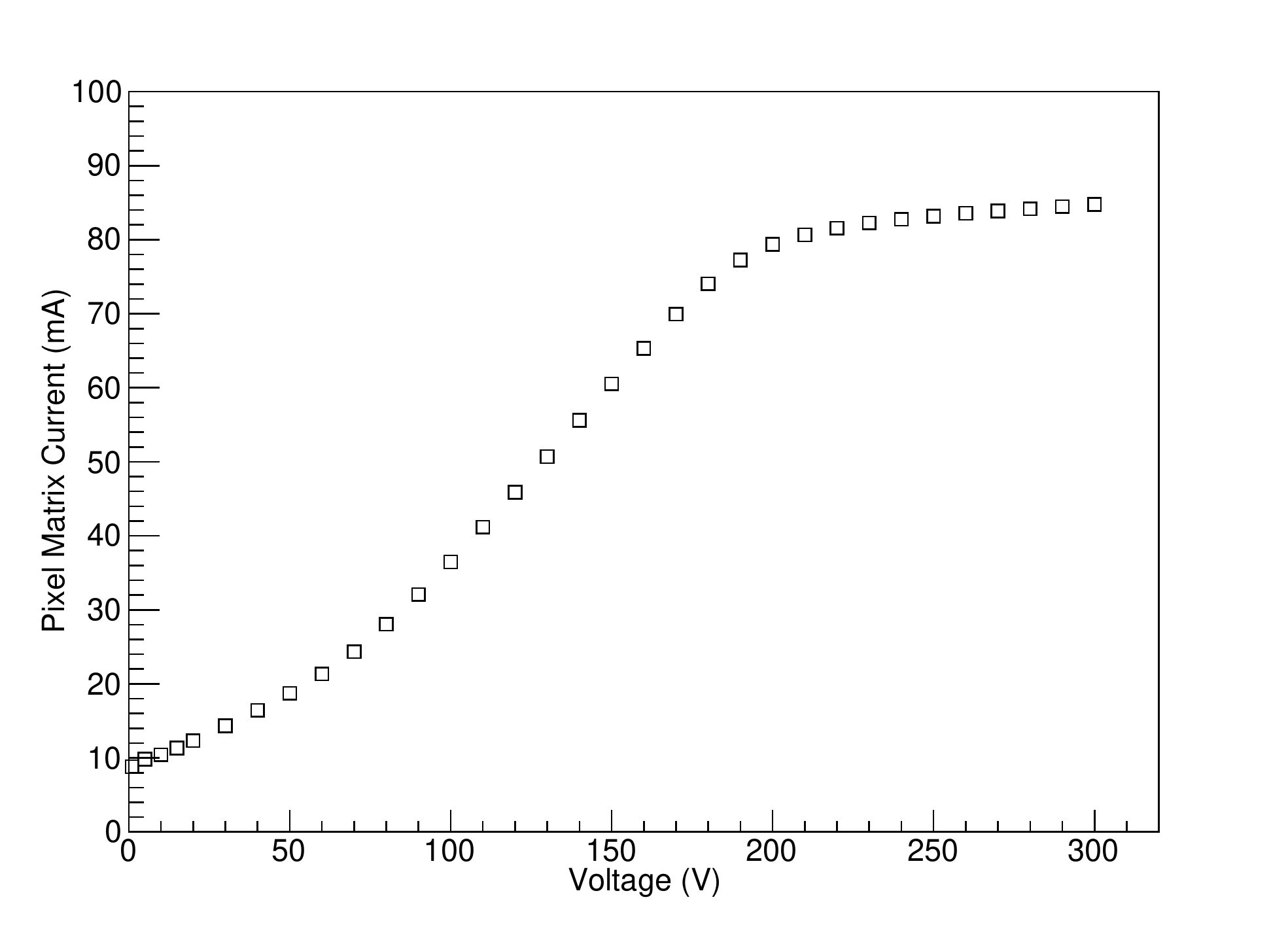}
\figcaption{\label{fig_response_voltage_iv} The signal amplitude of the seed pixel and summing over all the pixels (left). Leakage current of the pixel matrix as a function of the applied bias voltage (right).}
\end{center}

\subsection{Hit cluster}

The hit cluster size, critical for optimization of the readout scheme, was also studied. It can reveal the charge sharing among pixels as the carries diffuse/drift in the sensitive detector volume. Small clusters could be beneficial to enhance the SNR, while large clusters were possible to improve the spatial resolution with an appropriate clustering algorithm. Insteading of the 190 V full depletion voltage, a bias voltage of 20 V was applied to the substrate, which would then be less depleted. Figure~\ref{fig_20v_pixel_response} showed a typical hit cluster formed by the seed pixel and its four neighboring pixels. The signal amplitudes of the neighboring pixels were significantly lower than the directly illuminated seed pixel and dropped drastically as the distance to the seed pixel increased. Their contribution became negligible if they were several pixels away. This indicated that most of the stimulated charge in the sensitive detector volume would drift directly to the nearest charge collecting electrode. In addition, the average noise in each pixel was measured to be about 100 electrons. The lowest threshold should be set to be 5 to 6 times the ENC, i.e. 500 to 600 \(e\) accordingly. With a threshold of 600 \(e\), five pixels in total were fired. The cluster size was estimated to be 3 pixels or equivalently 60 \(\mu\)m. It should be noted that the hit cluster would spread less if real MIPs were used, because the divergence angle of the laser could lead to wider energy spread as the laser travelled deeper into the detector substrate and inevitably increase the hit cluster size.

The hit cluster size is critical for optimization of the readout scheme. It reveals the charge sharing among pixels when the carries diffuse/drift in the sensitive volume. Small clusters are beneficial to enhance SNR, while large clusters can improve the spatial resolution with an appropriate clustering algorithm. Insteading of the 190 V full depletion voltage, a bias voltage of 20 V was applied to the substrate, which would then be less depleted. Figure~\ref{fig_20v_pixel_response} shows a typical cluster formed by the seed pixel surrounded by four neighboring pixels, whose signal amplitudes were significantly lower than the directly illuminated seed pixel. Signals from other pixels droped drastically with increasing distance to the seed pixel. Their contributions became negligible when they were several pixels away. It indicates that most charge stimulated in the sensitive volume will drift directly to the nearest charge collecting electrode. The average noise in each pixel was measured to be about 100 electrons. The lowest threshold should be set to be 5 to 6 times the ENC, i.e. 500 to 600 \(e\). With the threshold set to 600 \(e\), five pixels in total were fired. The cluster size was estimated to be 3 pixels or equivalently 60 \(\mu\)m. It should be noted that hit cluster would spread less if real MIPs were used. The divergence angle of the laser pulse could lead to wider energy spread as the laser travels deeper into the sensor substrate and inevitably increases the hit cluster size.

\begin{center}
\includegraphics[width=5cm]{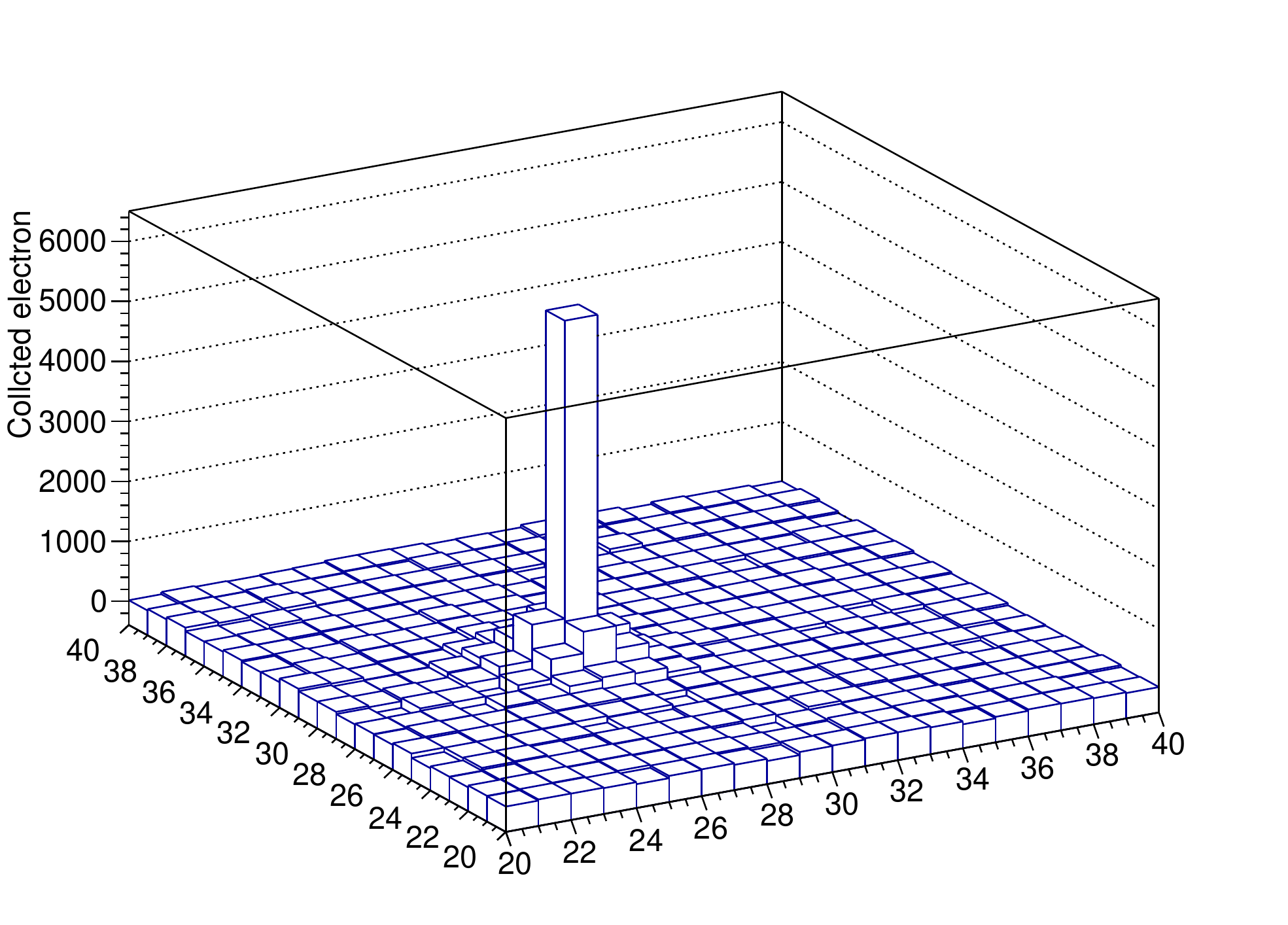}
\includegraphics[width=3.4cm]{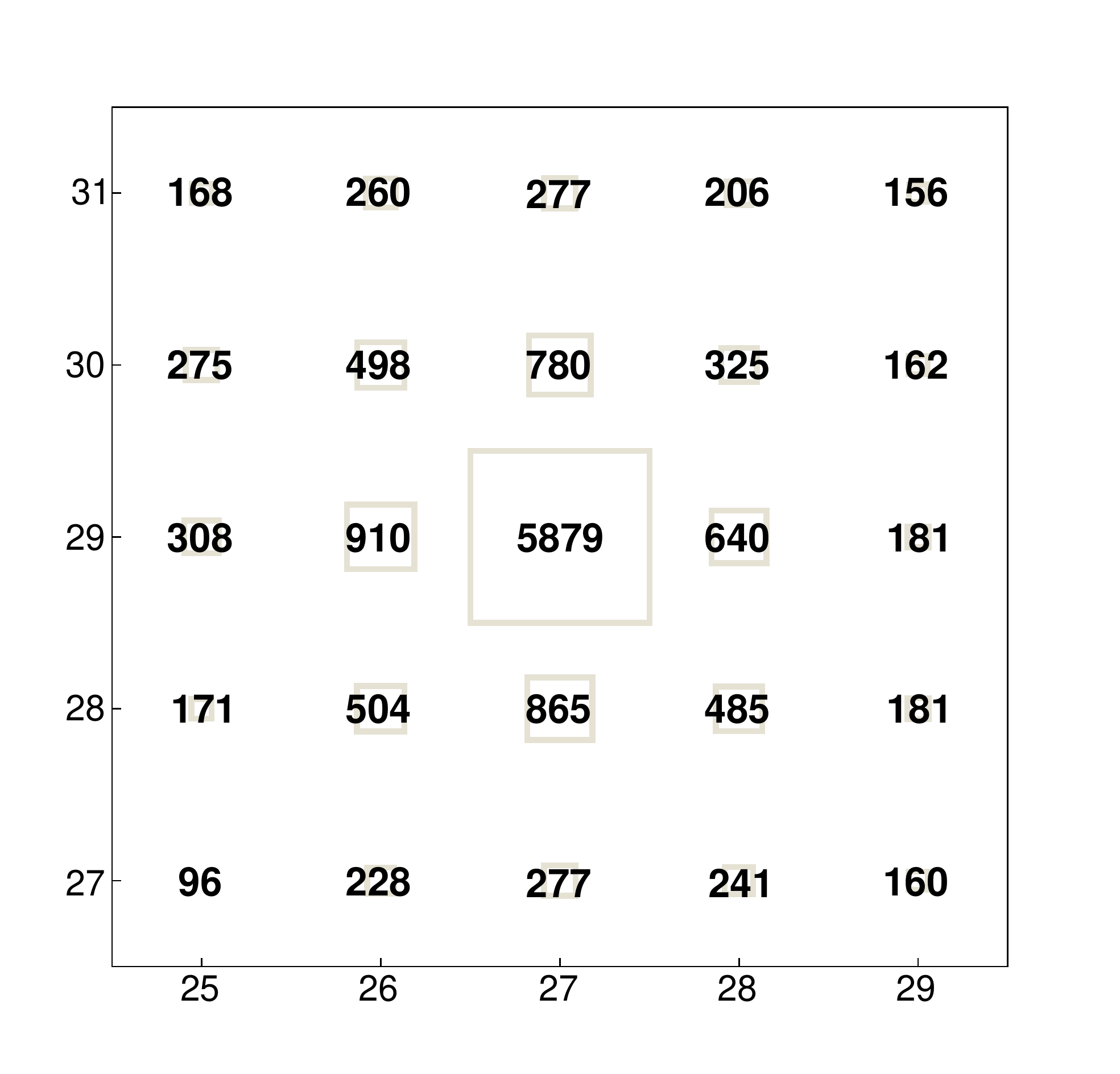}
\figcaption{\label{fig_20v_pixel_response} The amount of charge from a single hit collected by the seed pixel and its neighbors with 20 V bias voltage.}
\end{center}

It was also observed that the hit cluster size increased, when the higher bias voltage was applied. As shown in Fig.~\ref{fig_proportion}, the seed pixel collected a smaller fraction of the total stimulated charge as the bias voltage increased. The beam divergence and charge diffusion would result in wider hit clusters as the deeper depletion was reached.

\begin{center}
\includegraphics[width=7cm]{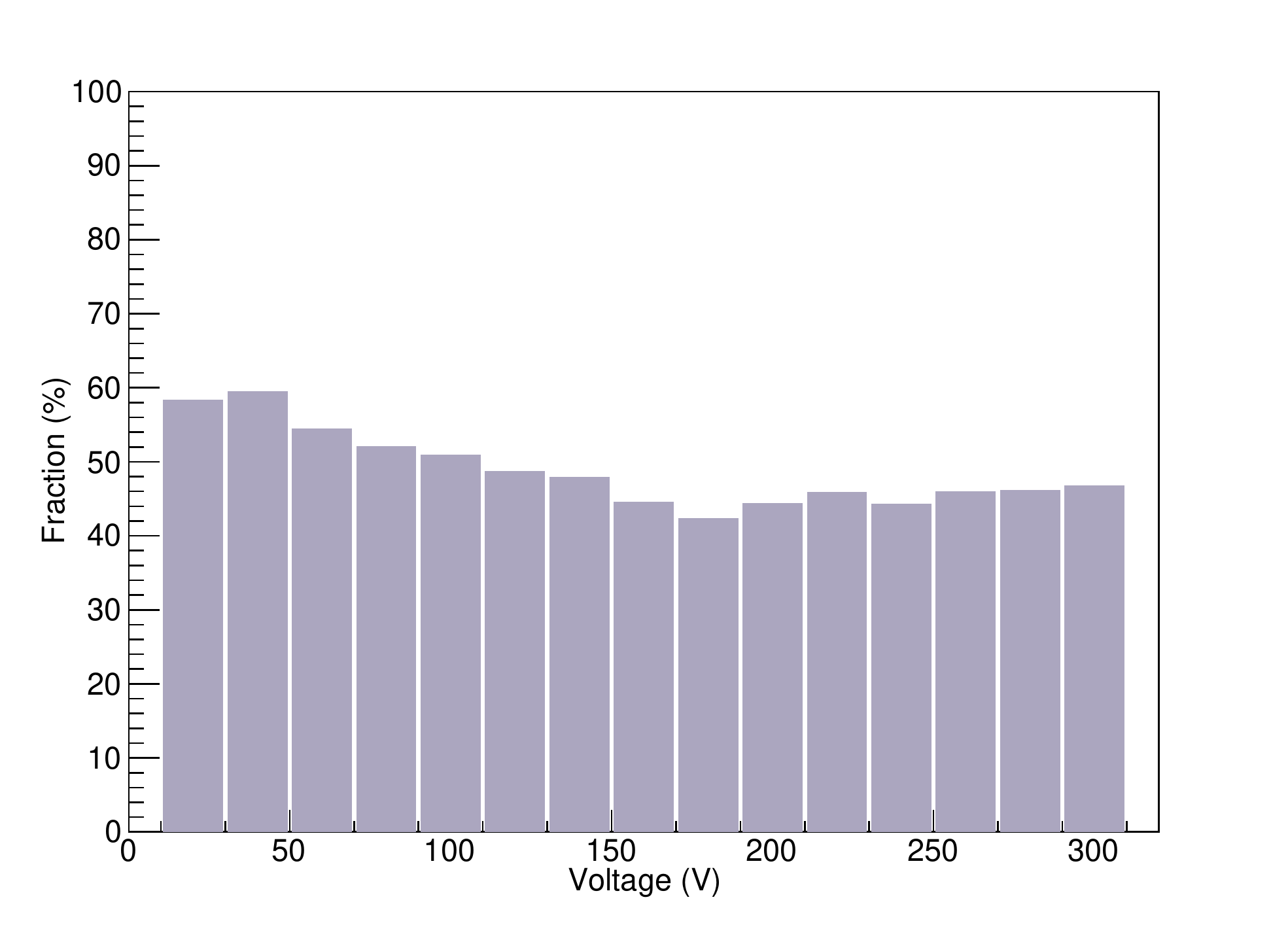}
\figcaption{\label{fig_proportion} Fraction of the seed pixel signal in total stimulated signal.}
\end{center}

\section{Conclusion}

A compact infrared laser system was developed to characterize a fine pitch SOI pixel detector. Laser pulses were tuned to a very short duration and highly focused to simulate MIP interaction with the silicon detector. The substrate depletion characteristics and hit cluster sizes of this SOI detector were studied. Compared to the classical IV tests, the test based on this laser system could determine more precisely the full depletion voltage. It was observed that most of the signal charge was collected by the directly illuminated pixel. The typical cluster size was estimated to be three pixels with the threshold set to six times of the noise level. In the future, this laser system will be integrated with other laboratory instruments, such as probe station and cooling tank, to further to enhance its in-house test capability.\\

\acknowledgments{The authors would like to thank the friendly SOIPIX researchers from KEK Insitute in Tsukuba (Japan) for their help in the SOI R\&D activities.}

\end{multicols}

\vspace{10mm}

\vspace{-1mm}
\centerline{\rule{80mm}{0.1pt}}
\vspace{2mm}

\begin{multicols}{2}

\end{multicols}

\clearpage

\end{CJK*}
\end{document}